\documentclass[a4paper,12pt]{article}
\usepackage[left=20mm,right=20mm,top=20mm,bottom=30mm]{geometry}
\usepackage{amsmath}
\usepackage{amssymb}
\usepackage[affil-it]{authblk}
\usepackage{graphicx}
\usepackage{subcaption}
\usepackage{hyperref}
\usepackage{microtype}
\usepackage{setspace}
\usepackage{tikz}
\usepackage{cite}
\usepackage{calc}
\usepackage{accents}
\usepackage{physics}
\usepackage{mathtools}

\usepackage{tikz}
\usetikzlibrary{arrows}
\usetikzlibrary{decorations.pathmorphing}
\usetikzlibrary{decorations.markings}
\usetikzlibrary{calc}

\tikzset{
  >=stealth',
  midarrow/.style={
    postaction={
      decorate,
      decoration={markings, mark=at position .55 with {\arrow{>}}}
    }
  },
  fermion/.style=midarrow,
  photon/.style={
    decorate,
    decoration={snake, amplitude=2pt, segment length=8pt}
  },
  boson/.style={
    decorate,
    decoration={snake, amplitude=2pt, segment length=8pt}
  },
  gluon/.style={
    decorate,
    decoration={coil, amplitude=4pt, segment length=5pt}
  },
  scalar/.style=densely dashed,
  arrowsnake/.style={
    preaction={photon, draw},
    postaction=midarrow
  }
}

\hypersetup{
  colorlinks,
  linkcolor={blue!50!black},
  citecolor={blue!50!black},
  urlcolor={blue!50!black}
}

\newcommand{\tikzvector}[6]{
  \draw [->]
    ([xshift=#3, yshift=#4]$#1!0.25!#2$)
    to node [#5] {#6}
    ([xshift=#3, yshift=#4]$#1!0.75!#2$)
}

\begin{document}

\title{Weak interaction corrections to muon pair production via the photon fusion at the LHC}

\author[1]{S.~I.~Godunov}
\author[1]{E.~K.~Karkaryan}
\author[1]{V.~A.~Novikov}
\author[2]{A.~N.~Rozanov}
\author[1]{M.~I.~Vysotsky}
\author[1]{E.~V.~Zhemchugov}

\affil[1]{
  \small I.~E.~Tamm Department of Theoretical Physics, Lebedev Physical
  Institute, \newline
  53 Leninskiy Prospekt, Moscow, 119991, Russia
}
 \affil[2]{
 \small Centre de Physique de Particules de Marseille (CPPM), Aux-Marseille
Universite, CNRS/IN2P3, \newline
 163 avenue de Luminy, case 902, Marseille, 13288, France}

\date{}

\maketitle

\begin{abstract}
  Analytical formulas describing the correction due to the $Z$ boson exchange to the cross section of the reaction $pp\rightarrow p\mu^+\mu^- X$ are presented. When the invariant mass of the produced muon pair $W\gtrsim 150~\text{GeV}$ and its total transverse momentum is large, the correction is of the order of 20\%.
\end{abstract}

\section{Introduction}

The ATLAS collaboration measured the cross section of muon pair production in ultraperipheral collisions (UPC) of protons at the Large Hadron Collider \cite{1708.04053}. In the case of UPC the lepton pair is produced in the $\gamma\gamma$ fusion and is accompanied by forward scattering of both protons. We calculated the corresponding cross section in \cite{2106.14842}. Our result agrees with the one obtained by the ATLAS collaboration within the experimental accuracy. The results obtained with the help of Monte Carlo simulations can be found in \cite{0803.0883, 1512.01178, 1410.2983, 1508.02718}. The $\mu^+\mu^-$ pair production occurs in $\gamma Z$ and $ZZ$ fusion as well. However, for the scattered proton to remain intact the square of the $4$-momentum of the emitted photon (or $Z$ boson) $Q^2$ should not considerably exceed $(200~\text{MeV})^2$. For larger $Q^2$ the cross section is suppressed by the elastic form factor \cite{1806.07238}. Thus the contribution of the diagram with the virtual $Z$ boson exchange is suppressed by the factor $Q^2/M^2_Z\sim 10^{-5}$ and can be safely omitted. But if only one of the protons remains intact while the second one dissolves producing a hadron jet then substitution of the photon emitted by the second proton by a $Z$ boson may lead to numerically noticeable corrections. This is so since the value of $Q^2$ is now bounded from above only by the invariant mass of the produced muon pair $W$ (for $Q^2>W^2$ the cross section of muon pair production is suppressed as a power of $W^2/Q^2$). In this way the contribution from the $Z$ boson exchange being proportional to $\qty[Q^2/(M^2_Z + Q^2)]^2$ can become noticeable for $W^2\gtrsim M^2_Z$. The ATLAS collaboration measured the fiducial cross section of the reaction of lepton pair ($e^+e^-$ or $\mu^+\mu^-$) production when one of the scattered protons is detected by the ATLAS Forward Proton Spectrometer \cite{2009.14537}. The other proton can remain intact or disintegrate. In the latter case the events were selected in which the momentum squared of the emitted photon did not exceed $(5~\text{GeV})^2$ (see our recent paper \cite{2207.07157} with formulas and numerical estimates of the cross sections measured in \cite{2009.14537}).

Dilepton production in proton-proton collisions through $\gamma\gamma$-fusion with the first proton scattered elastically while the second one produced a hadron jet was considered in our paper \cite{2112.01870}. Analytical formulas describing the cross section of a muon pair production were derived therein. In the present paper we calculate weak interaction corrections to the cross section obtained in \cite{2112.01870} which should be used for comparison of the Standard Model predictions with future experimental data. New Physics can manifest itself in interactions of muons at high energies. Thus accurate Standard Model predictions are needed to probe it. Tiny deviation of the measured value of muon magnetic moment from the SM result \cite{2104.03281, hep-ex/0602035} may indicate considerable deviation of muon pair production cross section at large invariant masses of muon pair $W \sim 100~\text{GeV}\divisionsymbol 1~\text{TeV}$. To find them the results of the present paper should be accounted for.

In Section~\ref{s:general} general formulas are derived. $Z$ boson exchange contributions are discussed in Section~\ref{s:correction}. Numerical results are presented in Section~\ref{s:numeric} and in Section~\ref{s:conclusion} we conclude.

\section{Muon pair production in $\gamma\gamma$ fusion}

\label{s:general}

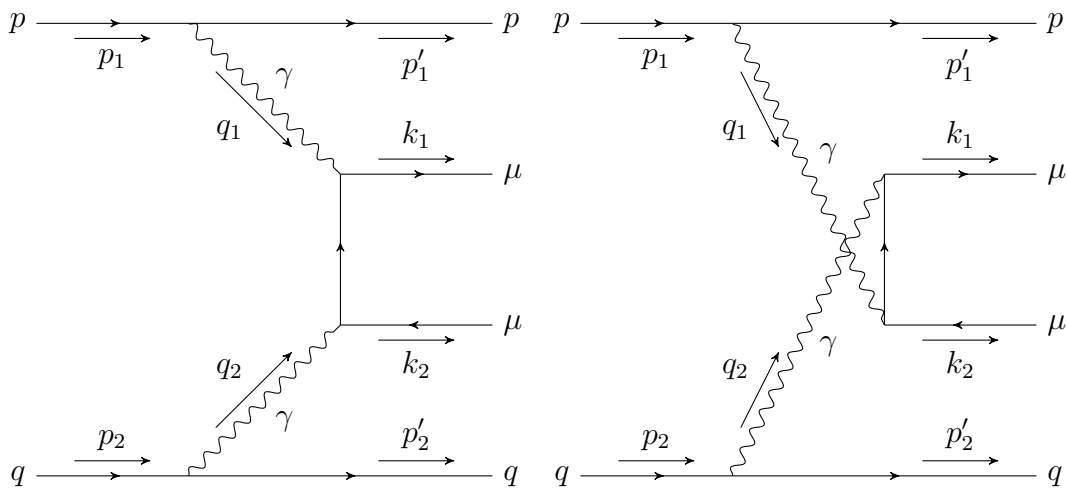
\begin{figure}[h]
  \centering
  \begin{tikzpicture}
    \coordinate (Pin)  at (-4,  3);
    \coordinate (Pout) at ( 2,  3);
    \coordinate (Qin)  at (-4, -3);
    \coordinate (Qout) at ( 2, -3);
    \coordinate (GP)   at (-2,  3);
    \coordinate (GQ)   at (-2, -3);
    \coordinate (GL1)  at ( 0,  1);
    \coordinate (GL2)  at ( 0, -1);
    \coordinate (Lout) at ( 2,  1);
    \coordinate (Lin)  at ( 2, -1);

    \draw [fermion] (Pin) node [left] {$p$} -- (GP);
    \draw [fermion] (GP) -- (Pout) node [right] {$p$};
    \draw [fermion] (Qin) node [left] {$q$} -- (GQ);
    \draw [fermion] (GQ) -- (Qout) node [right] {$q$};
    \draw [fermion] (Lin) node [right] {$\mu$} -- (GL2);
    \draw [fermion] (GL2) -- (GL1);
    \draw [fermion] (GL1) -- (Lout) node [right] {$\mu$};
    \draw [photon]  (GP) -- node [above right] {$\gamma$} (GL1);
    \draw [photon]  (GQ) -- node [below right] {$\gamma$} (GL2);

    \tikzvector{(Pin)}{(GP)}{0pt}{-2mm}{midway, below}{$p_1$};
    \tikzvector{(0, 3)}{(Pout)}{0pt}{-2mm}{midway, below}{$p'_1$};
    \tikzvector{(Qin)}{(GQ)}{0pt}{2mm}{midway, above}{$p_2$};
    \tikzvector{(0, -3)}{(Qout)}{0pt}{2mm}{midway, above}{$p'_2$};
    \tikzvector{(GP)}{(GL1)}{-1.4mm}{-1.4mm}{midway, below left}{$q_1$};
    \tikzvector{(GQ)}{(GL2)}{-1.4mm}{1.4mm}{midway, above left}{$q_2$};
    \tikzvector{(GL1)}{(Lout)}{0pt}{2mm}{midway, above}{$k_1$};
    \tikzvector{(GL2)}{(Lin)}{0pt}{-2mm}{midway, below}{$k_2$};
  \end{tikzpicture}
      \begin{tikzpicture}
          \coordinate (Pin)  at (-4,  3);
    \coordinate (Pout) at ( 2,  3);
    \coordinate (Qin)  at (-4, -3);
    \coordinate (Qout) at ( 2, -3);
    \coordinate (GP)   at (-2,  3);
    \coordinate (GQ)   at (-2, -3);
    \coordinate (GL1)  at ( 0,  1);
    \coordinate (GL2)  at ( 0, -1);
    \coordinate (Lout) at ( 2,  1);
    \coordinate (Lin)  at ( 2, -1);

    \draw [fermion] (Pin) node [left] {$p$} -- (GP);
    \draw [fermion] (GP) -- (Pout) node [right] {$p$};
    \draw [fermion] (Qin) node [left] {$q$} -- (GQ);
    \draw [fermion] (GQ) -- (Qout) node [right] {$q$};
    \draw [fermion] (Lin) node [right] {$\mu$} -- (GL2);
    \draw [fermion] (GL2) -- (GL1);
    \draw [fermion] (GL1) -- (Lout) node [right] {$\mu$};
    \draw [photon]  (GP) -- node [above right] {$\gamma$} (GL2);
    \draw [photon]  (GQ) -- node [below right] {$\gamma$} (GL1);

    \tikzvector{(Pin)}{(GP)}{0pt}{-2mm}{midway, below}{$p_1$};
    \tikzvector{(0, 3)}{(Pout)}{0pt}{-2mm}{midway, below}{$p'_1$};
    \tikzvector{(Qin)}{(GQ)}{0pt}{2mm}{midway, above}{$p_2$};
    \tikzvector{(0, -3)}{(Qout)}{0pt}{2mm}{midway, above}{$p'_2$};
    \tikzvector{(GP)}{(-1, 1)}{-1.4mm}{-1.4mm}{midway, below left}{$q_1$};
    \tikzvector{(GQ)}{(-1, -1)}{-1.4mm}{1.4mm}{midway, above left}{$q_2$};
    \tikzvector{(GL1)}{(Lout)}{0pt}{2mm}{midway, above}{$k_1$};
    \tikzvector{(GL2)}{(Lin)}{0pt}{-2mm}{midway, below}{$k_2$};
      \end{tikzpicture}
  \caption{Muon pair production in the photon fusion. The photon with momentum $q_1$ is radiated by the elastically scattered proton and the photon with momentum $q_2$ is radiated by the quark.}
  \label{f:2}
\end{figure}

Diagrams for muon pair production in ultraperipheral collisions of protons are shown in Fig.~\ref{f:2}.

The cross section for two-photon production can be expressed in terms of the amplitude $M_{\mu\alpha}$ of $\gamma\gamma\rightarrow\mu^+\mu^-$ transition as follows (see \cite{2207.07157} and the review of two-photon particle production \cite{budnev}):
\begin{align}
    \label{1}
    \dd\sigma_{pq\rightarrow p\mu^+\mu^-q} = & \, 2 \cdot \frac{Q^2_q\pqty{4\pi\alpha}^2}{\pqty{q^2_1}^2\pqty{q^2_2}^2} 
    \pqty{q^2_1\rho_{\mu\nu}^{\pqty{1}}\vphantom{q^2_2\rho_{\alpha\beta}^{\pqty{2}}}}
    \pqty{q^2_2\rho_{\alpha\beta}^{\pqty{2}}} 
    M_{\mu\alpha} M^{*}_{\nu\beta} \times \\ \nonumber
    & \times \frac{\pqty{2\pi}^4\delta^{\pqty{4}}\pqty{q_1+q_2-k_1-k_2} \, \dd\Gamma}{4\sqrt{\pqty{p_1p_2}^2 - m^4_p}}\frac{\dd^3p^{'}_1}{\pqty{2\pi}^3 2E^{'}_1} \frac{\dd^3p^{'}_2}{\pqty{2\pi}^3 2E^{'}_2}f_q(x, Q^2_2) \, \dd x,
\end{align}
where the leading factor 2 takes into account the symmetrical process where the other proton survives, $\alpha$ is the fine structure constant, $Q_q$ is the charge of quark $q$, $\rho^{(1)}_{\mu\nu}$ and $\rho^{(2)}_{\mu\nu}$ are the density matrices of the photons, $\dd\Gamma$ is the phase space of the muon pair, $m_p$ is the proton mass, $f_q\qty(x, Q^2_2)$ is the parton distribution function for quark $q$, $x$ is the fraction of the momentum of the disintegrating proton carried by the quark, $Q^2_2=-q^2_2$, $E_1'$ and $E_2'$ are the energies of the proton and the quark after the collision.

For the density matrix $\rho^{(1)}$ originating from the elastically scattered proton we have (see \mbox{Eqs.~(25-27)} from \cite{2207.07157}):
\begin{gather}
    \label{2}
    \rho^{(1)}_{\mu\nu} = -\qty(g_{\mu\nu} - \frac{q_{1\mu}q_{1\nu}}{q^2_1})G^2_M(Q^2_1) - \frac{\qty(2p_1 - q_1)_{\mu}\qty(2p_1-q_1)_{\nu}}{q^2_1}D\qty(Q^2_1),\\
    \nonumber
    D\qty(Q^2_1) = \frac{G^2_E\qty(Q^2_1) + \qty(Q^2_1/4m^2_p)G^2_M\qty(Q^2_1)}{1+Q^2_1/4m^2_p}.
\end{gather}
Here $Q_1^2=-q_1^2$, and $G_E\qty(Q^2_1)$, $G_M\qty(Q^2_1)$ are the Sachs form factors of the proton. For the latter we use the dipole approximation:
\begin{equation}
    G_E\qty(Q^2) = \frac{1}{\qty(1 + Q^2/\Lambda^2)^2},\quad G_M\qty(Q^2) = \frac{\mu_p}{\qty(1+Q^2/\Lambda^2)^2},\quad \Lambda^2 = \frac{12}{r^2_p}= 0.66~\text{GeV}^2,
\end{equation}
where $\mu_p = 2.79$ is the proton magnetic moment and $r_p = 0.84~\text{fm}$ is the proton charge radius \cite{form-factor}.

The density matrix of the photon emitted by the quark is 
\begin{equation}
    \label{4}
    \rho^{(2)}_{\alpha\beta} = -\frac{1}{2q^2_2} \Tr\{\hat{p}^{'}_2\gamma_{\alpha}\hat{p}_2\gamma_{\beta}\}
     = -\qty(g_{\alpha\beta} - \frac{q_{2\alpha}q_{2\beta}}{q^2_2}) - \frac{\qty(2p_2 - q_2)_{\alpha}\qty(2p_2-q_2)_{\beta}}{q^2_2}.
\end{equation}

It is convenient to consider the lepton pair production in the basis of virtual
photon helicity states. In the center of mass system (c.m.s.) of the colliding photons,
let $q_1 = (\tilde \omega_1, 0, 0, \tilde q)$, $q_2 = (\tilde \omega_2, 0, 0,
-\tilde q)$. The standard set of orthonormal 4-vectors orthogonal to $q_1$ and
$q_2$ is
\begin{equation}
    \label{5}
    \begin{aligned}
       e_1^+ &= \frac{1}{\sqrt{2}} (0, -1, -i, 0),
    &  e_1^- &= \frac{1}{\sqrt{2}} (0,  1, -i, 0),
    &  e_1^0 &= \frac{i}{\sqrt{-q_1^2}} (\tilde q, 0, 0, \tilde \omega_1),
    \\ e_2^+ &= \frac{1}{\sqrt{2}} (0,  1, -i, 0),
    &  e_2^- &= \frac{1}{\sqrt{2}} (0, -1, -i, 0),
    &  e_2^0 &= \frac{i}{\sqrt{-q_2^2}} (-\tilde q, 0, 0, \tilde \omega_2).
    \end{aligned}
\end{equation}
Due to the conservation of vector current, the covariant density matrices
$\rho_i^{\mu \nu}$ satisfy $q_1^\mu \rho_1^{\mu \nu} = q_2^\mu \rho_2^{\mu \nu}
= 0$. Thus, we can write
\begin{align}
    \label{7}
    \rho_i^{\mu \nu}
    &= \sum\limits_{a,b} \qty(e_i^{a \mu})^* e_i^{b \nu} \rho_i^{ab},
    \\
    \rho_i^{ab} &= (-1)^{a+b} e_i^{a \mu} \qty(e_i^{b \nu})^* \rho_i^{\mu \nu},
    \label{density-helicity}
\end{align}
where $a, b \in \{ \pm 1, 0 \}$, and $\rho_i^{ab}$ are the density matrices in the
helicity representation. The amplitudes of the lepton pair production in the
helicity basis $M_{ab}$ appear from the following equation:
\begin{align}
    \label{8}
    \rho_1^{\mu \nu} \rho_2^{\alpha \beta} M_{\mu \alpha} M^*_{\nu \beta}
    & = (-1)^{a + b + c + d} \rho_1^{ab} \rho_2^{cd} M_{ac} M^*_{bd} = \\ \nonumber
    & = \rho^{(1)}_{++}\rho^{(2)}_{++}|M_{++}|^2 + \rho^{(1)}_{++}\rho^{(2)}_{--}|M_{+-}|^2 + \rho^{(1)}_{++}\rho^{(2)}_{00}|M_{+0}|^2 + \\ \nonumber 
    & \hphantom{{}={} \rho^{(1)}_{++}\rho^{(2)}_{++}|M_{++}|^2} 
    + \rho^{(1)}_{--}\rho^{(2)}_{++}|M_{-+}|^2 
    + \rho^{(1)}_{--}\rho^{(2)}_{00}|M_{-0}|^2
    + \rho^{(1)}_{--}\rho^{(2)}_{--}|M_{--}|^2.
\end{align}
In this expression non-diagonal terms (those with $a \neq b$ or $c \neq d$) originated from the interference are omitted since their contributions cancel out when one integrates over azimuthal angles of the proton and the quark in the final state \cite{budnev, novikov}. The contribution of the longitudinally polarized photon emitted by the proton is neglected since it is proportional to $Q^2_1/W^2$, where $W$ is the invariant mass of the produced lepton pair. The reason is that due to the elastic form factors $Q^2_1$ is bounded by approximately $(200~\text{MeV})^2$ \cite{1806.07238} while $Z$ boson exchange corrections become noticeable at~$Q_1^2\sim M^2_Z$.

Matrix elements of the photon density matrices in the helicity representation for transverse polarizations were found in \cite{2207.07157} (see also \cite{budnev, novikov}). For the first photon we have:
\begin{equation}
    \label{9}
    \rho^{(1)}_{++} = \rho^{(1)}_{--} \approx D\qty(Q^2_1) \frac{2E^2 q^2_{1\perp}}{\omega_1^2 Q^2_1},
\end{equation}
where $E$ is the proton energy in the c.m.s.~of the colliding protons, $q_{1\perp}$ is the transversal momentum of the photon and $\omega_1$ is its energy in the same system. The function $D\qty(Q^2_1)$ is defined in \eqref{2}. In the following we will calculate the cross section of the muon pair production in the parton model. For the second photon which is emitted by the quark with the initial energy $xE$, $0<x<1$, we have:
\begin{equation}
    \label{10}
    \rho^{(2)}_{++} = \rho^{(2)}_{--} = \frac{2x^2E^2q^2_{2\perp}}{\omega^2_2 Q^2_2},\quad \rho^{(2)}_{00} = \frac{4x^2E^2q^2_{2\perp}}{\omega^2_2 Q^2_2}.
\end{equation}
Changing the integration variables from $\dd^3p^{'}_1 \dd^3 p^{'}_2$ to $\dd^3q_1 \dd^3q_2 = 2\pi (1/2)\dd q^2_{1\perp}\dd\omega_1 \cdot 2\pi (1/2) \dd q^2_{2\perp}\dd\omega_2 $ and substituting expressions (\ref{8}), (\ref{9}) and (\ref{10}) in (\ref{1}), we get:
\begingroup
\addtolength{\jot}{12pt}
\begin{align}
    \label{11}
    \dd\sigma_{pq \rightarrow p\mu^+\mu^-q} = & ~2Q^2_q (4\pi\alpha)^2 \frac{q_1q_2}{p_1p_2} 
    D\qty(Q^2_1) \frac{2E^2}{\omega_1^2} \frac{2x^2E^2}{\omega^2_2} \sigma_{\gamma\gamma^*\rightarrow\mu^+\mu^-}  \times \\ \nonumber 
    & \quad\times
    4\cdot \frac{q^2_{1\perp}}{Q^4_1} \frac{\frac{1}{2}\dd q^2_{1\perp}\dd\omega_1}{(2\pi)^2 2E}\cdot \frac{q^2_{2\perp}}{Q^4_2} \frac{\frac{1}{2}\dd q^2_{2\perp}\dd\omega_2}{(2\pi)^2 2Ex}f_q(x, Q^2_2)\dd x =\\ \nonumber 
    = & ~2\cdot \qty(\frac{\alpha}{\pi})^2 Q^2_q \frac{q_1q_2}{p_1p_2} \sigma_{\gamma\gamma^* \rightarrow \mu^+\mu^-} xE^2 \frac{D(Q^2_1) q^2_{1\perp}\dd q^2_{1\perp}}{Q^4_1} \frac{q^2_{2\perp}\dd q^2_{2\perp}}{Q^4_2} \frac{\dd\omega_1}{\omega^2_1} \frac{\dd\omega_2}{\omega^2_2}f_q(x, Q^2_2)\dd x,
\end{align}
\endgroup
where $\sigma_{\gamma\gamma^* \rightarrow \mu^+\mu^-}$ is the cross section of $\mu^+\mu^-$ production in $\gamma\gamma^*$ collision, $q_1q_2 = (W^2 - q^2_2)/2$ and $p_1p_2=2E^2x$. The differential cross section of muon pair production in the c.m.s.~of the photons is:
\begin{gather}
    \label{13}
    \dd\sigma_{\gamma\gamma^* \rightarrow \mu^+\mu^-} = \frac{\sum \overline{|M|^2} \, \dd\cos{\theta}}{32\pi W^2(1 + Q^2_2/W^2)}, \\ \nonumber
    \sum \overline{|M|^2} = \frac{1}{4} \Big[ |M_{++}|^2 + |M_{+-}|^2 + |M_{-+}|^2 + |M_{--}|^2 + 2|M_{+0}|^2 + 2|M_{-0}|^2 \Big],
\end{gather}
where $M_{ab}$ are the amplitudes of the process $\gamma\gamma^* \rightarrow \mu^+\mu^-$ in the helicity representation, $\theta$ is the scattering angle. It is convenient to express the corresponding cross section as the sum of the terms with the photon emitted by the quark polarized transversely and longitudinally: $\sigma_{\gamma\gamma^* \rightarrow \mu^+\mu^-} = \sigma_{TT} + \sigma_{TS}$, where according to Eq.~(E3) from \cite{budnev}
\begin{align}
    \label{14}
    \sigma_{TS} &\equiv \int \frac{1}{2} \qty[|M_{+0}|^2+|M_{-0}|^2]\frac{\dd\cos{\theta}}{32\pi W^2\qty(1+Q^2_2/W^2)} \approx \frac{16\pi\alpha^2W^2 Q^2_2}{\qty(W^2 + Q^2_2)^3},\\ \nonumber 
    \sigma_{TT} &\equiv \int \frac{1}{4}\qty[ |M_{++}|^2+|M_{+-}|^2+|M_{-+}|^2+|M_{--}|^2 ]\frac{\dd\cos{\theta}}{32\pi W^2\qty(1 + Q^2_2/W^2)} \approx  \\ \nonumber 
    &\approx 
    \frac{4\pi\alpha^2}{W^2}\qty[ \frac{1+Q^4_2/W^4}{\qty(1+Q^2_2/W^2)^3}\ln{\frac{W^2}{m^2}} - \frac{\qty(1-Q^2_2/W^2)^2}{\qty(1+Q^2_2/W^2)^3}],
\end{align}
where $m$ is the muon mass. Integration of \eqref{11}  with respect to $q^2_{1\perp}$ factors out the equivalent photon spectrum of the proton \cite[Eqs.~(4,~5)]{2207.07157}
\begin{align}
    \label{15}
    n_p\qty(\omega_1) &= \frac{\alpha}{\pi\omega_1}\int\limits^{\infty}_{0} \frac{D\qty(Q^2_1)q^2_{1\perp}\dd q^2_{1\perp}}{Q^4_1} = \\ \nonumber 
    &= 
    \frac{\alpha}{\pi\omega_1} \Bigg\{ \qty(1 + 4 u - (\mu_p^2 - 1) \frac{u}{v})\ln{\qty( 1 + \frac{1}{u})} - \frac{24 u^2 + 42 u + 17}{6 (u+1)^2}  - \\ \nonumber 
    &- \frac{\mu_p^2 - 1}{(v - 1)^3}\qty[\frac{1 + u / v}{v - 1}
              \ln \frac{u + v}{u + 1}
            - \frac{
                6u^2 (v^2 - 3v + 3) + 3u (3v^2 - 9v + 10) + 2v^2 - 7v + 11
              }{6 (u + 1)^2}] \Bigg\},
\end{align}
where $u = \qty(\omega_1/\Lambda\gamma)^2$, $v =\left(2 m_p/\Lambda \right)^2$, $q^2_1 \equiv -Q^2_1 \approx -q^2_{1\perp} - \omega^2_1/\gamma^2$ and $\gamma = E/m_p$.

It is convenient to change the integration variables in (\ref{11}) from the photon energies $\omega_1$ and $\omega_2$ to the square of the invariant mass of the produced pair $W^2 = 4\omega_1\omega_2 + q^2_2$ and its rapidity $y=1/2\ln{\omega_1/\omega_2}$: $\dd\omega_1\dd\omega_2 \dd q^2_{2\perp} = (1/4)\dd W^2\dd y\dd Q^2_2$. Taking into account that $q^2_2 \equiv -Q^2_2 \approx -q^2_{2\perp} - \omega^2_2/\gamma^2_q$, where $\gamma_q = E_q/m_q \approx 3xE/m_p = 3x\gamma$\footnote{Since the quark is bound within proton, we use the constituent quark mass $m_q = m_p/3 \approx 300~\text{MeV}$~\cite[60.1]{pdg}. In~\cite{2207.07157} it was checked that variation of $m_q$ from $200~\text{MeV}$ to $400~\text{MeV}$ changes the value of the cross section by few percents.}, we obtain:
\begin{equation}
    \label{16}
    \dd\sigma_{pq\rightarrow p\mu^+\mu^-q} = \frac{\alpha}{2\pi} Q^2_q n_{p}\qty(\omega_1) \sigma_{\gamma\gamma^* \rightarrow \mu^+\mu^-}\qty(W^2, Q^2_2) \frac{Q^2_2 - \qty(\omega_2/3x\gamma)^2}{\omega_2 Q^4_2}\dd W^2\dd y\dd Q^2_2 f_q\qty(x, Q^2_2)\dd x,
\end{equation}
where $\omega_1 = \sqrt{W^2+Q^2_2}\cdot e^y/2$, $\omega_2 = \sqrt{W^2+Q^2_2}\cdot e^{-y}/2$.

When we were considering this process in \cite{2207.07157}, we were working under experimental constraints that allowed us to use the approximation $Q^2_2 \ll W^2$. In that case the photon emitted by the quark is approximately real, the cross section for the photon fusion does not depend on $Q^2_2$, and we could factor out the function that we loosely interpreted as the equivalent photon spectrum of quark $q$:
\begin{equation}
    n_q(\omega)
  = \frac{2 \alpha Q_q^2}{\pi \omega}
    \int\limits_{\omega / E}^1 \mathrm{d} x
    \int\limits_0^{p_T^{\mu \mu}} \mathrm{d} q_{2 \perp}
    \, \frac{q_{2 \perp}^3}{Q_2^4} f_q\qty(x, Q_2^2),
\end{equation}
where $p^{\mu\mu}_T = 5~\text{GeV}$ is the experimental constraint on the muon system transversal momentum imposed in \cite{2207.07157}. We can use an analogous function here to simplify Eq.~\eqref{16} and to make it obvious how the violation of the equivalent photon approximation occurs in this problem:
\begin{align}
    n_q(\omega_2)
    &= \frac{\alpha Q_q^2}{\pi \omega_2}
    \int \mathrm{d} x
    \int \mathrm{d} Q^2_{2}
    \, \frac{q_{2 \perp}^2}{Q_2^4} f_q\qty(x, Q_2^2),\\
    \label{18}
    \frac{\dd n_q(\omega_2)}{\dd Q^2_2} &= \frac{\alpha Q^2_q}{\pi\omega_2}\int\limits_{x_{\rm min}}^{1} \frac{Q^2_2 - \qty(\omega_2/3x\gamma)^2}{Q^4_2}f_q\qty(x, Q^2_2)\dd x,
\end{align}
where 
\begin{equation}
    x_{\rm min} = \sqrt{\frac{W^2+Q^2_2}{s}} \cdot e^{-y} \cdot \max\qty{1, \frac{m_p}{3\sqrt{Q^2_2}}},
\end{equation}
$s = 4 E^2$ is the Mandelstam variable of the $pp \to p \mu^+ \mu^- q$ reaction. In this way from \eqref{16} we obtain:
\begin{equation}
    \label{19}
    \sigma_{pq \to p\mu^+\mu^-q} = \frac{1}{2} \int\limits_{\hat{W}^2}^s \dd W^2 \int\limits_{\frac{W^4}{36\gamma^2s}}^{s-W^2}\sigma_{\gamma\gamma^* \to \mu^+\mu^-}\qty(W^2, Q^2_2)\dd Q^2_2 \int\limits^{\frac{1}{2}\ln{\frac{s}{W^2+Q^2_2}}}_{\frac{1}{2}\ln{\qty(\frac{W^2+Q^2_2}{s}\cdot\max\qty(1,\frac{m_{p}^{2}}{9Q_{2}^{2}}))}} \hspace{-2em} n_p\qty(\omega_1)\frac{\dd n_q(\omega_2)}{\dd Q^2_2}\dd y,
\end{equation}
where we assume the experimental constraint on the invariant mass of the muon pair \mbox{$\hat W \gtrsim 10$}~GeV.

Finally, summing over valent $u$ and $d$ and sea quarks we get:
\begin{equation}
    \sigma_{pq \to p\mu^+\mu^-X} = \sum\limits_q \sigma_{pq \to p\mu^+\mu^-q}.
\end{equation}

\section{$Z$ boson exchange corrections}

\label{s:correction}

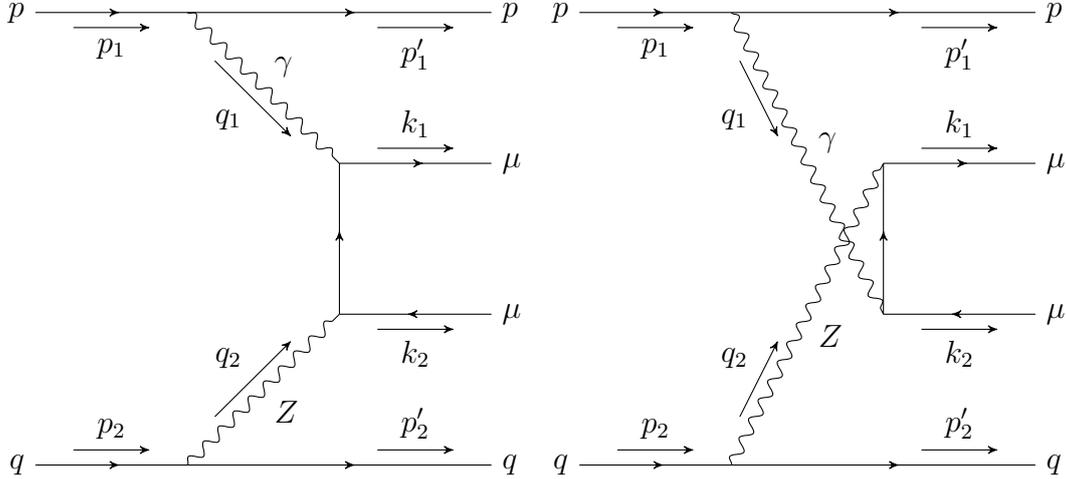
\begin{figure}[h]
  \centering
  \begin{tikzpicture}
    \coordinate (Pin)  at (-4,  3);
    \coordinate (Pout) at ( 2,  3);
    \coordinate (Qin)  at (-4, -3);
    \coordinate (Qout) at ( 2, -3);
    \coordinate (GP)   at (-2,  3);
    \coordinate (GQ)   at (-2, -3);
    \coordinate (GL1)  at ( 0,  1);
    \coordinate (GL2)  at ( 0, -1);
    \coordinate (Lout) at ( 2,  1);
    \coordinate (Lin)  at ( 2, -1);

    \draw [fermion] (Pin) node [left] {$p$} -- (GP);
    \draw [fermion] (GP) -- (Pout) node [right] {$p$};
    \draw [fermion] (Qin) node [left] {$q$} -- (GQ);
    \draw [fermion] (GQ) -- (Qout) node [right] {$q$};
    \draw [fermion] (Lin) node [right] {$\mu$} -- (GL2);
    \draw [fermion] (GL2) -- (GL1);
    \draw [fermion] (GL1) -- (Lout) node [right] {$\mu$};
    \draw [photon]  (GP) -- node [above right] {$\gamma$} (GL1);
    \draw [photon]  (GQ) -- node [below right] {$Z$} (GL2);

    \tikzvector{(Pin)}{(GP)}{0pt}{-2mm}{midway, below}{$p_1$};
    \tikzvector{(0, 3)}{(Pout)}{0pt}{-2mm}{midway, below}{$p'_1$};
    \tikzvector{(Qin)}{(GQ)}{0pt}{2mm}{midway, above}{$p_2$};
    \tikzvector{(0, -3)}{(Qout)}{0pt}{2mm}{midway, above}{$p'_2$};
    \tikzvector{(GP)}{(GL1)}{-1.4mm}{-1.4mm}{midway, below left}{$q_1$};
    \tikzvector{(GQ)}{(GL2)}{-1.4mm}{1.4mm}{midway, above left}{$q_2$};
    \tikzvector{(GL1)}{(Lout)}{0pt}{2mm}{midway, above}{$k_1$};
    \tikzvector{(GL2)}{(Lin)}{0pt}{-2mm}{midway, below}{$k_2$};
  \end{tikzpicture}
      \begin{tikzpicture}
          \coordinate (Pin)  at (-4,  3);
    \coordinate (Pout) at ( 2,  3);
    \coordinate (Qin)  at (-4, -3);
    \coordinate (Qout) at ( 2, -3);
    \coordinate (GP)   at (-2,  3);
    \coordinate (GQ)   at (-2, -3);
    \coordinate (GL1)  at ( 0,  1);
    \coordinate (GL2)  at ( 0, -1);
    \coordinate (Lout) at ( 2,  1);
    \coordinate (Lin)  at ( 2, -1);

    \draw [fermion] (Pin) node [left] {$p$} -- (GP);
    \draw [fermion] (GP) -- (Pout) node [right] {$p$};
    \draw [fermion] (Qin) node [left] {$q$} -- (GQ);
    \draw [fermion] (GQ) -- (Qout) node [right] {$q$};
    \draw [fermion] (Lin) node [right] {$\mu$} -- (GL2);
    \draw [fermion] (GL2) -- (GL1);
    \draw [fermion] (GL1) -- (Lout) node [right] {$\mu$};
    \draw [photon]  (GP) -- node [above right] {$\gamma$} (GL2);
    \draw [photon]  (GQ) -- node [below right] {$Z$} (GL1);

    \tikzvector{(Pin)}{(GP)}{0pt}{-2mm}{midway, below}{$p_1$};
    \tikzvector{(0, 3)}{(Pout)}{0pt}{-2mm}{midway, below}{$p'_1$};
    \tikzvector{(Qin)}{(GQ)}{0pt}{2mm}{midway, above}{$p_2$};
    \tikzvector{(0, -3)}{(Qout)}{0pt}{2mm}{midway, above}{$p'_2$};
    \tikzvector{(GP)}{(-1, 1)}{-1.4mm}{-1.4mm}{midway, below left}{$q_1$};
    \tikzvector{(GQ)}{(-1, -1)}{-1.4mm}{1.4mm}{midway, above left}{$q_2$};
    \tikzvector{(GL1)}{(Lout)}{0pt}{2mm}{midway, above}{$k_1$};
    \tikzvector{(GL2)}{(Lin)}{0pt}{-2mm}{midway, below}{$k_2$};
      \end{tikzpicture}
  \caption{Weak interaction corrections to muon pair production in semi-inclusive $pp\rightarrow p\mu^+\mu^- X$ reaction originating from the diagrams with $Z$ boson exchange.}
  \label{f:3}
\end{figure}
\newcommand{\dbtilde}[1]{\accentset{\approx}{#1}}

Corrections to the described by Fig.~\ref{f:2} and (\ref{1}) $\gamma\gamma$-fusion reaction come from the interference with the two diagrams shown in Fig.~\ref{f:3} and from the square of these diagrams. In both cases the expression for the density matrix $\rho^{(2)}_{\alpha\beta}$ given by (\ref{4}) should be modified in order to take into account the axial coupling of $Z$ boson to quarks. Let us designate these matrices $\Tilde{\rho}^{(2)}$ and $\dbtilde{\rho}^{(2)}$ correspondingly. Masses of quarks can be safely neglected making matrices $\Tilde{\rho}^{(2)}$ and $\dbtilde{\rho}^{(2)}$ transversal: $\Tilde{\rho}^{(2)}_{\alpha\beta}q_{2\beta} = \dbtilde{\rho}^{(2)}_{\alpha\beta}q_{2\beta} = 0$. That is why they can be expanded over the same $4$-vectors $e^a_{2\mu}$, presented in (\ref{5}). The coupling of $Z$ boson to quarks equals: 
\begin{gather}
    \Delta L_{qqZ} = \frac{e}{s_W c_W}\qty[ \frac{g^q_V}{2}\Bar{q}\gamma_{\alpha}q + \frac{g^q_A}{2}\Bar{q}\gamma_{\alpha}\gamma_5q ]Z_{\alpha},\quad s_W \equiv \sin{\theta_W}, \quad c_W \equiv \cos{\theta_W}, \\ \nonumber 
    g^q_V = T^q_3 - 2Q_q s^2_W,\quad g^q_A = T^q_3,
\end{gather}
where $e = \sqrt{4 \pi \alpha}$, $\theta_W$ is the electroweak mixing angle, $s_W^2 \approx 0.231 $~\cite{pdg}, and $T_3^q$ is the weak isospin of quark $q$, so for $\Tilde{\rho}_{\alpha\beta}^{(2)}$ and $\dbtilde{\rho}^{(2)}_{\alpha\beta}$ we obtain:
\begin{align}
    \Tilde{\rho}_{\alpha\beta}^{(2)} &= - \frac{1}{2q^2_2} \qty[ \frac{g^q_V}{2} \Tr\{\hat{p}^{'}_2 \gamma_{\alpha} \hat{p}_2 \gamma_{\beta}\} + \frac{g^q_A}{2} \Tr\{\hat{p}^{'}_2 \gamma_{\alpha} \hat{p}_2 \gamma_{\beta}\gamma_5\} ],
    \\
    \dbtilde{\rho}^{(2)}_{\alpha\beta} &= -\frac{1}{2q^2_2} \Tr\qty{\hat{p}^{'}_2 \qty(\frac{g^q_V}{2}\gamma_{\alpha} + \frac{g^q_A}{2}\gamma_{\alpha}\gamma_5 )\hat{p}_2 \qty( \frac{g^q_V}{2}\gamma_{\beta} + \frac{g^q_A}{2}\gamma_{\beta}\gamma_5 ) }.
\end{align}
Calculating matrix elements of $\Tilde{\rho}_{ab}^{(2)}$ in the helicity representation according to (\ref{density-helicity}), we get that the correction proportional to $g^q_A$ cancels in $\Tilde{\rho}_{00}^{(2)}$ and is proportional to $xE/\omega_2$ in $\Tilde{\rho}_{++}^{(2)}$ and $\Tilde{\rho}_{--}^{(2)}$, so it can be neglected when compared to the contribution of the correction proportional to $g_V^q$ which behaves as $(xE/\omega_2)^2$ (see (\ref{10})). Therefore:
\begin{equation}
    \Tilde{\rho}_{ab}^{(2)} \approx \frac{g^q_V}{2}\rho^{(2)}_{ab}.
\end{equation}
In the case of $\dbtilde{\rho}^{(2)}_{\alpha\beta}$ we observe similar suppression of terms originating from the terms proportional to the product $g^q_A\cdot g^q_V$. Neglecting them we obtain:
\begin{equation}
    \label{22}
    \dbtilde{\rho}^{(2)}_{ab} \approx \frac{\qty(g^q_V)^2 + \qty(g^q_A)^2}{4} \rho^{(2)}_{ab}.    
\end{equation}

The amplitude that corresponds to the sum of the diagrams in Fig.~\ref{f:2} and Fig.~\ref{f:3} equals:
\begin{equation}
    \label{25}
    \mathcal{A} = A_\mu \bar p' \gamma_\mu p,~
    A_\mu = \frac{eQ_q}{q^2_2}\bar q' \gamma_{\alpha} q M^{\gamma}_{\mu\alpha} + \frac{e}{s_Wc_W\qty(q^2_2 - M^2_Z)} \bar q' \qty[\frac{g^q_V}{2}\gamma_{\alpha} + \frac{g^q_A}{2}\gamma_{\alpha}\gamma_5]q M^Z_{\mu\alpha}.
\end{equation}
For the $\gamma\gamma \to \mu^+\mu^-$ and $\gamma Z \to \mu^+\mu^-$ amplitudes (see Fig.~\ref{f:3.1}) we have:
\begingroup
\addtolength{\jot}{12pt}
\begin{align}
    \label{26}
    M^{\gamma}_{\mu\alpha} &= e^2\qty[\bar \mu \gamma_{\mu} \frac{1}{\hat k_1 - \hat q_1 - m}\gamma_{\alpha}\mu + \bar \mu \gamma_{\alpha} \frac{1}{\hat q_1 - \hat k_2 - m}\gamma_{\mu}\mu],\\ 
    {~}\vspace{1em}
    \label{27}
    M^Z_{\mu\alpha} &= \frac{Q_{\mu}e^2}{s_W c_W} \qty{\frac{g_V^{\mu}}{2}\qty[\bar \mu \gamma_{\mu} \frac{1}{\hat k_1 - \hat q_1 - m}\gamma_{\alpha}\mu + \bar \mu \gamma_{\alpha} \frac{1}{\hat q_1 - \hat k_2 - m}\gamma_{\mu}\mu] + \frac{g^{\mu}_A}{2}\qty[\gamma_{\alpha} \to \gamma_{\alpha}\gamma_5]} = \\ \nonumber 
    &= \frac{Q_{\mu}}{s_W c_W}\frac{g^{\mu}_V}{2}M^{\gamma}_{\mu\alpha} + \frac{Q_{\mu}e^2}{s_W c_W}\frac{g^{\mu}_A}{2}\qty[\bar \mu \gamma_{\mu} \frac{1}{\hat k_1 - \hat q_1 - m}\gamma_{\alpha}\gamma_5\mu + \bar \mu \gamma_{\alpha}\gamma_5 \frac{1}{\hat q_1 - \hat k_2 - m}\gamma_{\mu}\mu].
\end{align}
\endgroup
Substituting \eqref{26} and \eqref{27} into \eqref{25} and separating the term proportional to $M^{\gamma}_{\mu\alpha}$, we obtain:

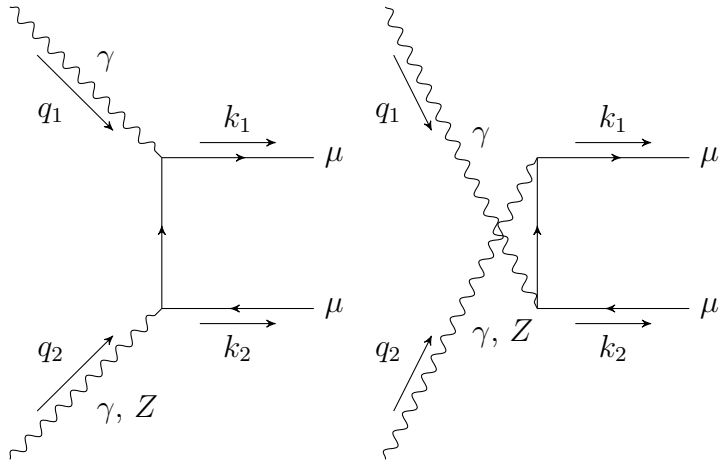
\begin{figure}[h]
  \centering
  \begin{tikzpicture}
    \coordinate (GP)   at (-2,  3);
    \coordinate (GQ)   at (-2, -3);
    \coordinate (GL1)  at ( 0,  1);
    \coordinate (GL2)  at ( 0, -1);
    \coordinate (Lout) at ( 2,  1);
    \coordinate (Lin)  at ( 2, -1);

    \draw [fermion] (Lin) node [right] {$\mu$} -- (GL2);
    \draw [fermion] (GL2) -- (GL1);
    \draw [fermion] (GL1) -- (Lout) node [right] {$\mu$};
    \draw [photon]  (GP) -- node [above right] {$\gamma$} (GL1);
    \draw [photon]  (GQ) -- node [below right] {$\gamma$, $Z$} (GL2);

    \tikzvector{(GP)}{(GL1)}{-1.4mm}{-1.4mm}{midway, below left}{$q_1$};
    \tikzvector{(GQ)}{(GL2)}{-1.4mm}{1.4mm}{midway, above left}{$q_2$};
    \tikzvector{(GL1)}{(Lout)}{0pt}{2mm}{midway, above}{$k_1$};
    \tikzvector{(GL2)}{(Lin)}{0pt}{-2mm}{midway, below}{$k_2$};
  \end{tikzpicture}
      \begin{tikzpicture}
    \coordinate (GP)   at (-2,  3);
    \coordinate (GQ)   at (-2, -3);
    \coordinate (GL1)  at ( 0,  1);
    \coordinate (GL2)  at ( 0, -1);
    \coordinate (Lout) at ( 2,  1);
    \coordinate (Lin)  at ( 2, -1);

    \draw [fermion] (Lin) node [right] {$\mu$} -- (GL2);
    \draw [fermion] (GL2) -- (GL1);
    \draw [fermion] (GL1) -- (Lout) node [right] {$\mu$};
    \draw [photon]  (GP) -- node [above right] {$\gamma$} (GL2);
    \draw [photon]  (GQ) -- node [below right] {$\gamma$, $Z$} (GL1);

    \tikzvector{(GP)}{(-1, 1)}{-1.4mm}{-1.4mm}{midway, below left}{$q_1$};
    \tikzvector{(GQ)}{(-1, -1)}{-1.4mm}{1.4mm}{midway, above left}{$q_2$};
    \tikzvector{(GL1)}{(Lout)}{0pt}{2mm}{midway, above}{$k_1$};
    \tikzvector{(GL2)}{(Lin)}{0pt}{-2mm}{midway, below}{$k_2$};
      \end{tikzpicture}
  \caption{$\mu^+\mu^-$ pair production in $\gamma\gamma$ (or $\gamma$Z) fusion.}
  \label{f:3.1}
\end{figure}

\begin{align}\label{28}
    A_\mu =& \qty{\bar q' \gamma_{\alpha} q \qty( \frac{e Q_q}{q^2_2} + \frac{g^{\mu}_V}{2}\frac{e Q_{\mu}}{\qty(s_W c_W)^2 \qty(q^2_2 - M^2_Z)} \frac{g^q_V}{2} ) +  \bar q' \gamma_{\alpha}\gamma_5 q\frac{g^{\mu}_V}{2}\frac{e Q_{\mu}}{\qty(s_W c_W)^2\qty(q^2_2 - M^2_Z)}\frac{g^q_A}{2}} \times \\ \nonumber
    &\hspace{50mm} \times e^2\qty[\bar \mu \gamma_{\mu} \frac{1}{\hat k_1 - \hat q_1 - m}\gamma_{\alpha}\mu + \bar \mu \gamma_{\alpha} \frac{1}{\hat q_1 - \hat k_2 - m}\gamma_{\mu}\mu] + \\ \nonumber
    &+\bar q' \qty(\frac{g^q_V}{2}\gamma_{\alpha} + \frac{g^q_A}{2}\gamma_{\alpha}\gamma_5)q \frac{g^{\mu}_A}{2}\frac{e Q_{\mu}}{\qty(s_W c_W) \qty(q^2_2 - M^2_Z)} \frac{e^2}{s_W c_W}\times \\ \nonumber 
    &\hspace{50mm}\times \qty[\bar \mu \gamma_{\mu} \frac{1}{\hat k_1 - \hat q_1 - m}\gamma_{\alpha}\gamma_5\mu + \bar \mu \gamma_{\alpha}\gamma_5 \frac{1}{\hat q_1 - \hat k_2 - m}\gamma_{\mu}\mu].
\end{align}

The following two statements are proved in Appendix: (1) the amplitudes in the square brackets which describe the vector and axial couplings to muons do not interfere and (2) the square of the amplitude with the axial coupling equals that with the vector coupling (we are working in $W\gg m$ domain). In this way for the square of the amplitude we get:
\begin{align}
    \abs{\mathcal{A}}^2 \equiv \varkappa \abs{\mathcal{A}_{\gamma\gamma}}^2,\hspace{10mm}
   & \varkappa\left(Q_{2}^{2}\right) = 1 +
    2\cdot\frac{g^{\mu}_{V}}{Q_{\mu}}\cdot\frac{g^{q}_{V}}{Q_{q}}\cdot\lambda
    +\frac{\left(g^{\mu}_{V}\right)^{2}+\left(g^{\mu}_{A}\right)^{2}}{Q_{\mu}^{2}}\cdot
    \frac{\left(g^{q}_{V}\right)^{2}+\left(g^{q}_{A}\right)^{2}}{Q_{q}^{2}}\cdot\lambda^{2},\\
  & \hspace{10mm} \lambda\equiv\frac{1}{\left(2s_{W}c_{W}\right)^{2}\left(1+M_{Z}^{2}/Q_{2}^{2}\right)},
\end{align}
where $\mathcal{A}_{\gamma\gamma}$ is the amplitude for the reaction via photon fusion only.

Thus we get that the squares of the helicity amplitudes describing the $\gamma\gamma^* \to \mu^+\mu^-$ reaction presented in \eqref{13} should be multiplied by the same factor $\varkappa$ in order to take into account the diagrams with the $Z$ boson exchange. It is convenient to calculate the helicity amplitudes $M^{\gamma}$ in the muons c.m.s.

The sum of the squares of the amplitudes of the processes with the longitudinal polarization of the vector boson (Z or photon) emitted by the quark is:
\begin{equation}
    \label{30}
    \abs{\tilde{M}_{+0}}^2 + \abs{\tilde{M}_{-0}}^2 = \qty(4\pi\alpha)^2 \frac{32Q^2_2}{W^2\qty(1+Q^2_2/W^2)^2}\varkappa,
\end{equation}
and the corresponding cross section is
\begin{equation}
    \tilde \sigma_{TS} = \frac{16\pi\alpha^2 Q^2_2}{W^4  \qty(1 + Q^2_2/W^2)^3}\varkappa.
\end{equation}
It equals zero for $Q^2_2 = 0$, as it should be. For the contribution of the transversal amplitudes with opposite helicities we obtain:
\begin{equation}
    \abs{\tilde{M}_{+-}}^2 + \abs{\tilde{M}_{-+}}^2 = \qty(4\pi\alpha)^2\frac{4\sin^2\theta\qty(1+\cos^2\theta)}{\qty(1+Q^2_2/W^2)^2}\qty[\frac{1}{1-v\cos{\theta}} + \frac{1}{1+v\cos{\theta}}]^2 \varkappa.
\end{equation}
Since the sum of helicities of the produced leptons equals one, while that of the annihilating vector bosons equals two, the production at $\theta=0,~\pi$ is forbidden by the helicity conservation. The factor $\sin\theta$ takes care of it. For the contribution to the cross section we obtain: 
\begin{equation}
    \label{32}
    \tilde \sigma_{+-} + \tilde \sigma_{-+} = \frac{4\pi\alpha^2}{W^2\qty(1+Q^2_2/W^2)^3}\qty[\ln\qty(\frac{W^2}{m^2}) - 2]\varkappa,
\end{equation}
where $m$ is a lepton mass.

Performing straightforward calculations, for the square of the amplitude induced by photons with helicities $e^+_{1,2}$ we obtain:
\begin{align}
    \label{33}
    \abs{M^{\gamma}_{++}}^2 = 4\qty(4\pi\alpha)^2 \left\{\sin^2{\theta} \frac{Q^4_2/W^4}{\qty(1+Q^2_2/W^2)^2} \qty[ \frac{1}{\qty(1-v\cos{\theta})^2} + \frac{1}{\qty(1+v\cos{\theta})^2} ] + \vphantom{\frac{1-v^2}{\qty(1-v\cos{\theta})^2} + \frac{1-v^2}{\qty(1+v\cos{\theta})^2}} \right. \hspace{2em}\\ \nonumber \left.
    \vphantom{\sin^2{\theta} \frac{Q^4_2/W^4}{\qty(1+Q^2_2/W^2)^2} \qty[ \frac{1}{\qty(1-v\cos{\theta})^2} + \frac{1}{\qty(1+v\cos{\theta})^2} ]} + \frac{1-v^2}{\qty(1-v\cos{\theta})^2} + \frac{1-v^2}{\qty(1+v\cos{\theta})^2}\right\},
\end{align}
where $v$ is the muon velocity in the c.m.s., $v^2 = 1 - 4m^2/W^2$. Helicity of the initial state of the two photons equals zero while for massless muons helicity of the $\mu^+\mu^-$ pair equals $\pm1$. That is why the amplitude $M_{++}$ should be zero at $\theta = 0, \pi$ when the helicity is conserved (massless muons). The factor $\sin^2{\theta}$ multiplying the first term in the curly braces takes care of this. In the case of muon (or antimuon) spin flip the final state will have zero helicity, and muons production at $\theta=0, \pi$ is allowed. The two last terms in the curly braces are responsible for the forward and backward muon production. Though their numerators are proportional to $m^2$, they make finite contribution to the cross section in the limit $m\rightarrow 0$ due to the denominators which come from muon propagators. Integrating these terms over $\theta$, we obtain a finite in the limit $m/W \rightarrow 0$ contribution to the cross section which comes from the $\theta \sim m/W$ and $\theta \sim \pi - m/W$ domains. This phenomena is the essence of the chiral anomaly\cite{Adler, Jackiv}: even in the limit $m\rightarrow 0$ production of muons at $\theta =0, \pi$ is still allowed.

The contribution to the cross section of muon pair production is the following:
\begin{align}
    \label{34}
    \tilde \sigma_{++} + \tilde \sigma_{--} &= \int \frac{\frac{1}{4}\qty(|\tilde M_{++}|^2 + |\tilde M_{--}|^2)}{32 \pi W^2 \qty(1+ Q^2_2/W^2)}\dd\cos{\theta} = \\ \nonumber &= \frac{4\pi\alpha^2}{W^2\qty(1+Q^2_2/W^2)} \qty[ \frac{Q^4_2/W^4}{\qty(1+Q^2_2/W^2)^2}\qty( \ln{\frac{W^2}{m^2}} - 2 ) +1 ]\varkappa
\end{align}
and $\tilde \sigma_{TT} = \tilde \sigma_{+-} + \tilde \sigma_{-+} + \tilde \sigma_{++} + \tilde \sigma_{--} $, where $\tilde \sigma_{+-}+\tilde \sigma_{-+}$ is given by (\ref{32}). Let us note that even in the case of the collision of real photons ($Q^2_2=0$) the cross section $\tilde \sigma_{++} + \tilde \sigma_{--}$ is non-zero. As a final result we get:
\begin{equation}
    \label{35}
    \tilde\sigma_{TT} = \frac{4\pi\alpha^2}{W^2}\qty[ \frac{1+Q^4_2/W^4}{\qty(1+Q^2_2/W^2)^3}\ln{\frac{W^2}{m^2}} - \frac{\qty(1-Q^2_2/W^2)^2}{\qty(1+Q^2_2/W^2)^3} ]\varkappa.
\end{equation}

We have discussed $\gamma\gamma$ and $\gamma Z$ contributions. Let us note that the $ZZ$ contribution should be very small for the process under consideration since it is suppressed by $Q_1^2/M_Z^2\lesssim 10^{-5}$.

\section{Numerical results}

\label{s:numeric}

Replacing in \eqref{19}
$\sigma_{\gamma\gamma^* \to \mu^+\mu^-}\qty(W^2, Q_2^2)$ with
$\tilde \sigma_{TS} + \tilde \sigma_{TT}$ given in \eqref{30} and
\eqref{35} we obtain an expression for the differential cross section
of a muon pair production which takes into account the $Z$ boson exchange:
\begin{align}
  \frac{\dd\sigma_{pp\rightarrow p\mu^+\mu^-X}}{\dd W} =
  &\frac{4\alpha W}{\pi} \sum_q
    Q^2_q \int\limits^{s}_{\frac{W^4}{36\gamma^2s}}
    \frac{\qty[\tilde\sigma_{TS}\qty(W^2, Q^2_2) + \tilde\sigma_{TT}\qty(W^2,
    Q^2_2)]}{W^2+Q^2_2}\dd Q^2_2 \times \\ \nonumber
  & \times
    \int\limits^{\frac{1}{2}\ln{\frac{s}{W^2+Q^2_2}}}_{\frac{1}{2}\ln{\qty(\frac{W^2+Q^2_2}{s}\cdot\max\qty(1,\frac{m_{p}^{2}}{9Q_{2}^{2}}))}}
    \omega_1 n_p(\omega_1) \dd y \int\limits^1_{x_{\rm min}}
    \frac{Q^2_2-\qty(\omega_2/3x\gamma)^2}{Q^4_2}f_q\qty(x, Q^2_2)\dd x,
\end{align}
where we used the equality $\omega_2 = (W^2 + Q_2^2) / 4 \omega_1$.

For better convergence of the numerical integration it is convenient
to change the order of integrals:
\begin{align}
  \label{eq:main}
  \frac{\dd\sigma_{pp\rightarrow p\mu^+\mu^-X}}{\dd W} =
  & \frac{4\alpha W}{\pi} \sum_q Q^2_q
    \int\limits^{s}_{\frac{W^4}{36\gamma^2s}}
    \frac{\sigma_{\gamma\gamma^* \to \mu^+\mu^-}\qty(W^2,
    Q_2^2)}{\qty(W^2+Q^2_2)Q^4_2}\cdot\varkappa\left(Q_{2}^{2}\right)\cdot
    \dd Q^2_2 \times \\ \nonumber
  & \times
    \int\limits^1_{\frac{W^{2}+Q_{2}^{2}}{s}\cdot\max\qty(1,\frac{m_{p}}{3\sqrt{Q_{2}^{2}}})}\hspace{-2em}\dd x f_q(x, Q^2_2)
    \int\limits^{\frac{1}{2}\ln{\frac{s}{W^2+Q^2_2}}}_{\frac{1}{2}\ln{\qty(\frac{W^2+Q^2_2}{x^2s}\cdot\max\qty(1,\frac{m_{p}^{2}}{9Q_{2}^{2}}))}}
    \hspace{-2em}\omega_1 n_p(\omega_1) \qty[Q^2_2-\qty(\omega_2/3x\gamma)^2] \dd y,\\
  & \varkappa\left(Q_{2}^{2}\right) = 1 +
    2\cdot\frac{g^{\mu}_{V}}{Q_{\mu}}\cdot\frac{g^{q}_{V}}{Q_{q}}\cdot\lambda
    +\frac{\left(g^{\mu}_{V}\right)^{2}+\left(g^{\mu}_{A}\right)^{2}}{Q_{\mu}^{2}}\cdot
    \frac{\left(g^{q}_{V}\right)^{2}+\left(g^{q}_{A}\right)^{2}}{Q_{q}^{2}}\cdot\lambda^{2},\\
  & \hspace{10mm} \lambda\equiv\frac{1}{\left(2s_{W}c_{W}\right)^{2}\left(1+M_{Z}^{2}/Q_{2}^{2}\right)}.
\end{align}
In this formula we explicitly separated
$\varkappa\left(Q_{2}^{2}\right)$ since it is the only source of weak
interaction corrections in~\eqref{eq:main}. For all charged fermions
in the Standard Model $g_{V}$ and $g_{A}$ have the same sign as their
charge $Q$, so the weak interaction correction is positive.

Let us make a couple of comments regarding formula~\eqref{eq:main} and
its accuracy. The lower integration limit on $Q_{2}^{2}$
in~\eqref{eq:main} is much smaller than
$\left(1~\text{GeV}\right)^{2}$. So we should explain if it is
consistent with the parton approximation. This lower limit comes from
kinematics and does not take into account other requirements. However,
there is a soft cut at small $Q_{2}^{2}$ in the parton distribution
functions provided by LHAPDF~\cite{1412.7420} so we decided not to
change this limit in~\eqref{eq:main}. Numerically, the region
$Q_{2}^{2}<\left(1~\text{GeV}\right)^{2}$ gives $\sim 15\%$ of the
differential cross section for $W=20~\text{GeV}$ and even less for
larger $W$. We must stress that it does not affect the absolute value
of the weak interaction correction since
$\varkappa\left(Q_{2}^{2}\right)=1$ for
$Q_{2}^{2}\lesssim \left(1~\text{GeV}\right)^{2}$ with very good
accuracy.

We should also note that the photon flux of the disintegrating proton is
not fully described by the parton approximation. Resonance phenomena and
other effects can increase the cross section by
10--15\%~\cite{1708.01256,2107.02535,1510.00294}. However, the
corresponding region $Q_{2}^{2}\sim \left(1~\text{GeV}\right)^{2}$ is
not relevant for the weak interaction correction, so we do not take
these effects into account.

\begin{figure}[p]
  \centering
  \includegraphics[width=5.75in]{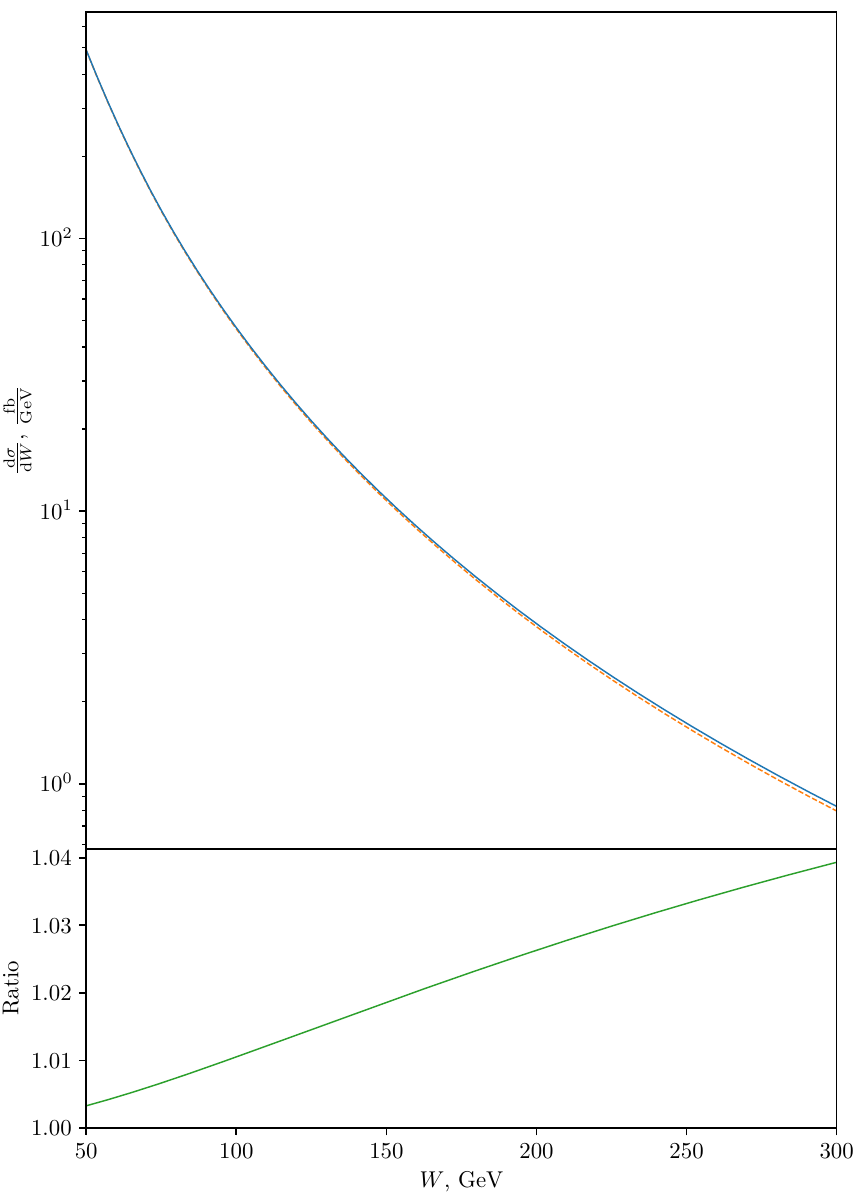}
  \caption{\textit{Upper plot:} differential cross section of the photon fusion only (orange dashed line) and with the weak interaction correction added (blue solid line). \textit{Lower plot:} their ratio. }
    \label{fig:full}
\end{figure}

Results of our numerical calculation are shown in Fig.~\ref{fig:full}. We
see that the weak interaction correction does not give a noticeable
increase in the cross section. The reason for that is clear: all
scales of $Q_{2}^{2}$ in \eqref{eq:main} are equally important while
for the weak interaction correction only the domain
$M_{Z}^{2}\lesssim Q_{2}^{2}\lesssim W^{2}$ is relevant.

\begin{figure}[p]
  \centering
  \begin{subfigure}[b]{5in}
    \includegraphics[width=\textwidth]{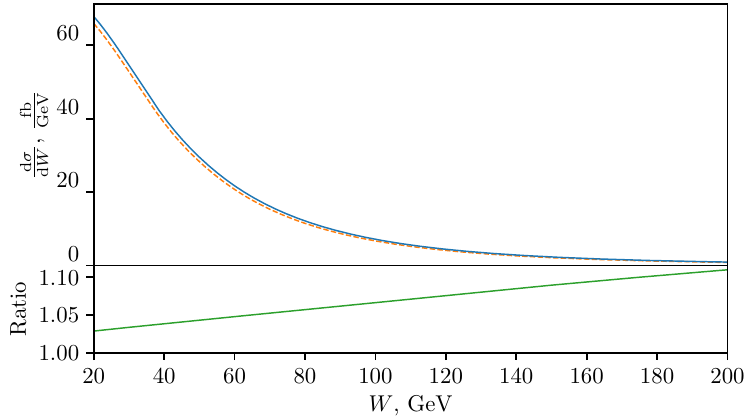}
    \caption{$\hat{Q}_{2}=30~\text{GeV}$.}
    \label{fig:dsigma_30}
  \end{subfigure}\\
  \begin{subfigure}[b]{5in}
    \includegraphics[width=\textwidth]{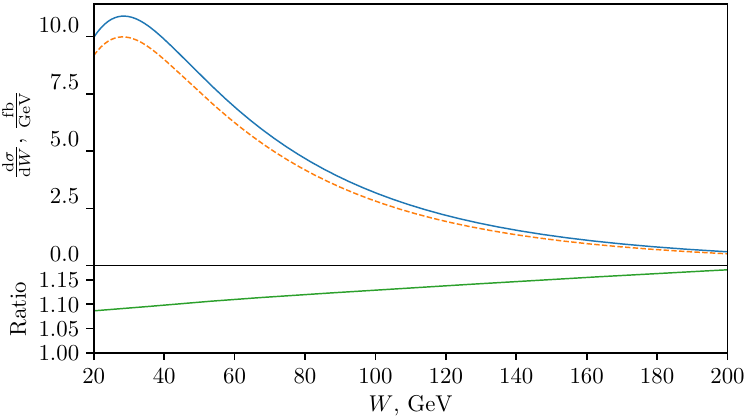}
    \caption{$\hat{Q}_{2}=50~\text{GeV}$.}
    \label{fig:dsigma_50}
  \end{subfigure}\\
  \begin{subfigure}[b]{5in}
    \includegraphics[width=\textwidth]{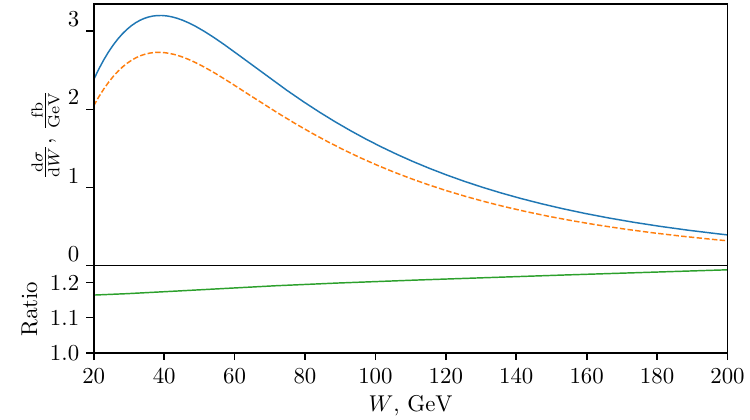}
    \caption{$\hat{Q}_{2}=70~\text{GeV}$.}
    \label{fig:dsigma_70}
  \end{subfigure}
  \vspace{-7pt}
  \caption{Differential cross sections for different lower limits on $Q_2^2$. Styles and colors
    of the lines are the same as in Fig.~\ref{fig:full}.}
  \label{fig:Q2min}
\end{figure}

The weak interaction correction is more pronounced when we set the lower
limit on $Q_{2}^{2}$, $Q_{2}^{2}>\hat{Q}_{2}^{2}$, closer to the
electroweak scale, $M_{Z}^{2}$. For
$\hat{Q}_{2}\gg 1~\text{GeV}$ it means a cut on
the total transverse momentum of the produced pair,
$p_{T}^{\mu\mu}>\hat{Q}_{2}$. Such a selection is
both theoretically and experimentally clean. From the theory side we
have no complications due to low-$Q^{2}$ physics. From the experiment
side this selection means high-$p_{T}$ events with two distinguishable
muons (at least for $W\gtrsim \hat{Q}_2$ since the cross section is additionally suppressed for
$Q_{2}^{2}>W^{2}$, see~\eqref{eq:main}). The results for
$\hat{Q}_{2}=30,~50,~70~\text{GeV}$ are shown in
Fig.~\ref{fig:Q2min}. We see that the weak interaction correction
 reaches 20 \%.

In the limit $Q_2^2\gg M_Z^2$ the function $\varkappa\qty(Q_2^2)$ reaches its maximum value:
\begin{align}
  \label{eq:kappa_max}
  \varkappa\left(Q_{2}^{2}\right) & \approx 1 +
    2\cdot\frac{g^{\mu}_{V}}{Q_{\mu}}\cdot\frac{g^{q}_{V}}{Q_{q}}
    \cdot\frac{1}{\left(2s_{W}c_{W}\right)^{2}}
    +\frac{\left(g^{\mu}_{V}\right)^{2}+\left(g^{\mu}_{A}\right)^{2}}{Q_{\mu}^{2}}\cdot
    \frac{\left(g^{q}_{V}\right)^{2}+\left(g^{q}_{A}\right)^{2}}{Q_{q}^{2}}
    \cdot\frac{1}{\left(2s_{W}c_{W}\right)^{4}}\approx \\ \nonumber
  & \approx
  \left\{
  \begin{aligned}
      1.35 & \quad\text{for}\quad q=u,~\bar u,~c,~\bar c;\\
      2.76 & \quad\text{for}\quad q=d,~\bar d,~s,~\bar s,~b,~\bar b.
  \end{aligned}
  \right.
\end{align}
However, to get an idea of what the correction in principle might be, one should keep in mind that $u$- and $c$- quarks and antiquarks give $\sim 80~\%$ of the cross section in photon--photon fusion.

\section{Conclusions}

\label{s:conclusion}

We have calculated the weak interaction correction to the cross
section of lepton pair production in semi-exclusive
process. Beforehand it was not obvious if this correction is
significant or not. It turned out that it gives few percent increase
of the production cross section. However, if we set the lower limit on the
net transverse momentum of the produced pair, the correction goes up
and can reach~20~\%.

Numerical calculations were performed with the help of \texttt{libepa}~\cite{libepa}. For the parton distribution functions we use \texttt{MMHT2014nnlo68cl}~\cite{1412.3989} provided by LHAPDF~\cite{1412.7420}.

We are supported by RSF grant No. 19-12-00123-$\Pi$.

\section*{Appendix}
\setcounter{equation}{0}
\renewcommand{\theequation}{A\arabic{equation}}
Let us prove that the square of the expression in the square brackets in \eqref{28} which contains the product of the vector currents $\gamma_{\mu}\cdot \gamma_{\alpha}$ equals that which contains the product of the vector and axial currents $\gamma_{\mu}\cdot \gamma_{\alpha}\gamma_5$. This statement is evident for massless muons, $m=0$. Thus taking into account the nonzero mass it is convenient to commute the muons propagators with the vector current~$\gamma_{\mu}$: 
\begin{align}
    \bar{\mu}_1 \gamma_{\mu}\qty( \hat k_1 - \hat q_1 + m ) = \bar \mu_1\qty(2k_{1\mu} - \gamma_{\mu}\hat q_1),\\
    \qty( \hat q_1 - \hat k_2 + m )\gamma_{\mu}\bar{\mu}_2 = \qty(\hat q_1 \gamma_{\mu} - 2k_{2\mu})\bar \mu_2.
\end{align}
Then the mass remains only in the muons polarization density matrices from which two types of terms can give nonzero contribution to the cross section: $\sim\hat k_1 \hat k_2$ and $\sim m^2$. The terms proportional to the product of the  momenta are the same for the vector-vector currents product and the vector-axial currents product, while those proportional to $m^2$ have the opposite signs. Let us prove that contributions of the terms proportional to $m^2$ are negligible.

In order to produce finite contributions in the limit $m^2/W^2\to 0$ they should contain squares of the muon propagators $\sim 1/\qty(k_1q_1)^2$ or $1/\qty(k_2q_1)^2$. The terms proportional to $m^2/\qty(k_1q_1)^2$ are multiplied by $\Tr\qty[\qty(2k_{1\mu} - \gamma_{\mu}\hat q_1)\gamma_{\alpha}\gamma_{\beta}\qty(2k_{1\nu} - \hat q_1 \gamma_{\nu})]$. For the amplitudes $M_{\pm 0}$ we get: 
\begin{equation}
    \Tr\qty[\qty(2k_{1\mu} - \gamma_{\mu}\hat q_1)\hat e^0_2 \hat e^0_2\qty(2k_{1\nu} - \hat q_1 \gamma_{\nu})] = 16 k_{1\mu}k_{1\nu},
\end{equation}
where we take into account that $q_{1\mu}M_{\mu\alpha} = q_{1\nu}M_{\nu\beta}=0$ and neglect $q^2_1$. Since $k_{1\mu}e^{\pm}_{1\mu}\sim \sin\theta$, even for the scattering angles $\theta\sim 0$ or $\pi$ we obtain negligible contribution into differential cross section $\sim m^2/W^2 \cdot \sin^2\theta/\qty(1\pm v\cos\theta)^2$. Analogous consideration is valid for the term proportional to~$1/\qty(k_2q_1)^2$.

In the case of transversal polarization of the photon or the $Z$ boson emitted by the quark, the trace discussed above contains a sum of terms proportional to $q^2_1$, $\qty(q_1e^{\pm}_2) = 0$, $\qty(q_1 e^{\pm}_1) =0$ and $\qty(k_{1\mu}e^{\pm}_{1\mu})\qty(k_{1\nu}e^{\pm}_{1\nu})\sim\sin^2\theta$. In this way we demonstrate that the terms proportional to $m^2$ can be safely neglected which proves that the square of the amplitude which contains axial coupling equals that which contains only vector couplings.

It remains to prove that the amplitudes in the square brackets in \eqref{28} do not interfere. Let us multiply the expressions in the square brackets by the vector boson polarization vectors \eqref{5} and introduce notations $M^V_{ik}$ and $M^A_{ik}$ for the terms in the first and the second square brackets correspondingly. Using the identity $\Tr\qty[\gamma_A\gamma_B...\gamma_C\gamma_D] = \Tr\qty[\gamma_D\gamma_C...\gamma_B\gamma_A]$ one can check that the following relations are valid:
\begin{align}
    M^V_{++}\qty[M^A_{++}]^* + M^A_{--}\qty[M^V_{--}]^* &= M^A_{++}\qty[M^V_{++}]^* + M^V_{--}\qty[M^A_{--}]^* = 0,
    \\
    M^V_{+0}\qty[M^A_{+0}]^* + M^A_{-0}\qty[M^V_{-0}]^* &= M^A_{+0}\qty[M^V_{+0}]^* + M^V_{-0}\qty[M^A_{-0}]^* = 0,
    \\
    M^V_{+-}\qty[M^A_{+-}]^* + M^A_{-+}\qty[M^V_{-+}]^* &= M^A_{+-}\qty[M^V_{+-}]^* + M^V_{-+}\qty[M^A_{-+}]^* = 0.
\end{align}
These relations are the consequence of the different behaviour of the vector and axial helicity amplitudes under the $P$-parity inversion: for example, $M^V_{+-} \xrightarrow{P} -M^V_{-+}$ and $M^A_{+-} \xrightarrow{P} M^A_{-+}$. It follows that in the cross section the interference terms cancel.

\newcommand{\arxiv}[1]{arXiv:\nolinebreak[3]\href{http://arxiv.org/abs/#1}{#1}}

\end{document}